\documentclass[aps,prb,twocolumn,amsmath,amssymb,showpacs,superscriptaddress]{revtex4}

\usepackage{bm}
\usepackage{graphicx}

\begin{document}


\title{Nonlinear magnetotransport in a dc-current-biased graphene}

\author{C. M. Wang}
\email[]{cmwangsjtu@gmail.com} \affiliation{School of Physics and
Electrical Engineering, Anyang Normal University, Anyang 455000,
China}
\author{X. L. Lei}
\affiliation{Department of Physics, Shanghai Jiaotong University,
1954 Huashan Road, Shanghai 200030, China}

\date{\today}

\begin{abstract}
A balance-equation scheme is developed to investigate the
magnetotransport in a dc-current-biased graphene. We
examine the Shubnikov-de Haas oscillation under a nonzero bias
current. With an increase in the current density, the oscillatory differential
resistivity exhibits phase inversion, in agreement with recent
experimental observation. In the presence of surface optical phonons,
a second phase inversion may occur at higher dc bias,
due to the reduced influence of electron-heating and the enhanced direct effect of current
on differential magnetoresistivity.
We also predict the appearance of current-induced magnetoresistance oscillation
 in suspended graphene at lower magnetic fields and larger current densities.
 For the graphene mobility currently available ($\approx 20\,{\rm m^2/Vs}$),
 the oscillatory behavior may be somewhat altered by magnetophonon resonance arising from
 intrinsic acoustic phonon under finite bias current condition.
\end{abstract}

\pacs{75.47.-m, 72.80.Vp, 73.50.Fq}

\maketitle

\section{introduction}
Since its isolation for the first time in
2004,\cite{novoselov2004electric} graphene, a two-dimensional (2D)
single-layer of carbon atoms, has attracted an explosion of
interest\cite{geim2007rise,sarma2010electronic,goerbig2011electronic}
due to both its fundamental physics and its potential technological
applications. In contrast to ordinary semiconductors, the
application of a strong perpendicular magnetic field on pristine
graphene results in an energetic quantization proportional to the
square root of external magnetic field with the existence of a true
zero-energy sharing equally by electrons and holes. As a result,
magnetotransport in graphene may exhibits unusual properties. For
example, the unique quantum Hall effect in graphene showing
half-integer Hall
plateaus,\cite{zhang2005experimental,novoselov2007room} has become
the experimental evidence of massless linear-energy fermionic
excitation.

Similarly, the resistivity minima of Shubnikov-de Hass oscillation
(SdHO) of graphene appears when filling factor equals
$4\left(n+\tfrac{1}{2}\right)$ with $n$ an integer. Recently, Tan
{\it et al.}\cite{tan2011shubnikov} found that in addition to the
damp of oscillation due to elevated carrier temperature, a phase
inversion of the differential magnetoresistivity occurs under dc
bias in graphene with relatively low zero-field mobility, i.e. SdHO
maxima (minima) invert to minima (maxima). They attributed the
observed interesting phenomenon to the elevated electron
temperature. The dominant energy dissipation they referred arises
from the diffusion of hot carriers to electrodes. However, when a
graphene is on a polar substrate, inelastic carrier scattering with
surface optical phonons (SO phonons) is important and offers an
intrinsic energy-dissipation
mechanism.\cite{fratini2008substrate,li2010influence,zhu2009carrier}
This notable phase-inversion effect has also been observed
experimentally in usual two-dimensional electron gas (2DEG) with
high mobility.\cite{kalmanovitz2008warming} So far, a microscopic
theoretical analysis including carrier--phonon scattering effect on
dc-current-induced phase inversion of SdHO has still been lacking
even for parabolic energy-band system.

The magnetoresistance oscillation directly induced by a dc current,
periodic in current density and in inverse magnetic field, is
another noteworthy nonlinear transport phenomenon, which was first
observed a decade ago in conventional
2DEG.\cite{yang2002zener,bykov2005effect} The effect is ascribed to
the Zener tunneling between Hall-field-tilted Landau levels due to
short-range (SR) impurity
scattering.\cite{yang2002zener,zhang2007magnetotransport} Later, by
including inelastic phonon scattering, a microscopic
balance-equation scheme has been constructed, conveniently
accounting for this current-induced nonlinear transport phenomenon
with considering the electron heating.\cite{lei2007current} So far,
however, investigations of nonlinear magnetoresistance oscillation
have been carried out only for high-mobility 2DEGs with parabolic
energy
dispersion.\cite{yang2002zener,zhang2007magnetotransport,bykov2005effect,lei2007current,zhang2007effect,vavilov2007}
Owing to the absence of much of extrinsic impurities and SO phonons,
the suspended graphene can achieve relatively high
mobility\cite{bolotin2008ultrahigh,bolotin2008temperature} that
Landau levels can be well resolved even in quite a weak magnetic
field. Hence, it is expected that this kind of nonlinear
magnetoresistance oscillation could been observed in suspended
graphene. Therefore, an efficient scheme capable of dealing with
magnetotransport in graphene under an external current bias is in
sore need.

The balance-equation approach, which is especially suitable to deal
with current-controlled nonlinear transport, was established based
on the separation of the center-of-mass motion from the relative
carrier motion in parabolic energy-band
systems.\cite{lei1985gsf,lei1985tdb,cai1985,lei1987nonlinear,lei2008balance}
It turns out that this scheme can be applied to systems with linear
energy dispersion.\cite{wang2012linear} In this paper, we will
generalize this scheme to graphene subject to a magnetic field and a
finite dc current. The paper is organized as follows. In
Sec.\,\ref{bef}, the force- and energy-balance equations are derived
for graphene in the presence of normal magnetic field and external
dc current. The effect of a finite dc bias on the SdHO in a graphene
on a SiO$_2$ substrate is investigated in Sec.\,\ref{sdho}. The
current-control magnetoresistance oscillation in a suspended
graphene is discussed in Sec.\,\ref{CIMO}. A summary is given in
Sec.\,\ref{summary}. The derivation of energy-balance equation is
presented in Appendix.

\section{Balance-Equation Formulation}\label{bef}

We consider a single layer graphene  in the $x$-$y$ plane under the
influence of a uniform magnetic field ${\bm B}=B\hat{z}$ along the
$z$ direction and a dc electric field ${\bm E}=(E_x,E_y)$ applied in
the layer plane. The carriers having enough density near the $K$ or
$K'$ points in the graphene, are interacting with each other,
coupled with lattice vibrations of the graphene as well as the oxide
interface, and scattered by randomly located disorders. The
Hamiltonian of this system consists of an carrier part $\mathcal
H_{\rm e}$, a phonon part $\mathcal H_{\rm ph}$, and
carrier--impurity and carrier--phonon interactions $\mathcal H_{\rm
ei}$ and $\mathcal H_{\rm ep}$:
\begin{equation}\label{}
\mathcal H=\mathcal H_{\rm e}+\mathcal H_{\rm ph}+\mathcal H_{\rm
ei}+\mathcal H_{\rm ep}.
\end{equation}
Here, the carrier Hamiltonian can be written as
\begin{equation}\label{helectron}
\mathcal H_{\rm e}=\sum_{j,\alpha}\left[v_{\rm
F}(\pi_{j}^x\sigma_j^x+c_\alpha\pi_{j}^y\sigma_j^y)+ e\bm
r_j\cdot\bm E \right] ,
\end{equation}
where ${\bm r}_j=(x_{j},y_{j})$, ${\bm p}_j=(p_{jx},p_{jy})$, ${\bm \pi}_j\equiv
\bm p_j+e\bm A(\bm r_j)=(\pi_j^x, \pi_j^y)$,   and ${\bm
\sigma}_j=(\sigma_{j}^x,\sigma_{j}^y,\sigma_{j}^z)$,
 stand, respectively, for the coordinate, momentum, canonical momentum, and Pauli
 operators of the $j$th carrier in the pseudospin
space formed by the A and B sublattices; $c_\alpha$ is a
valley-related coefficient equaling $+1$ or $-1$ for carrier in
$\alpha=K$ or $K'$ valley; $\bm A(\bm r)=(-By,0)$ is the vector
potential of the magnetic field in the Landau gauge; $v_{\rm
F}=1.1\times10^6\,{\rm m/s}$ is the Fermi velocity. The forms of
$\mathcal H_{\rm ei}$ and $\mathcal H_{\rm ep}$ are similar to those
given in Refs.\,\onlinecite{lei1987nonlinear} and
\onlinecite{lei2008balance}, without intervalley transition of
carriers.

In the frame work of balance equation
approach,\cite{lei1985tdb,lei1985gsf,cai1985} we introduce the 2D
center-of-mass (c.m.) momentum and coordinate $\bm
P_\alpha=\sum_{j\in\alpha}\bm p_j$ and $\bm
R_\alpha=N^{-1}_\alpha\sum_{j\in\alpha}\bm r_j$, and the
relative-carrier momenta and coordinates $\bm p'_j=\bm p_j-\bm
P_\alpha/N_\alpha$ and $\bm r_j'=\bm r_j-\bm R_\alpha$ for carriers
in the $\alpha$ valley having carrier number $N_{\alpha}$, to write
the Hamiltonian $\mathcal H_{\rm e}$ into the sum of a
single-particle c.m. part $\mathcal H_{\rm cm}=\sum_\alpha \mathcal
H_{\rm cm}^{\alpha}$
 and a many-particle relative-carrier part $\mathcal H_{\rm er}=\sum_\alpha \mathcal H_{\rm er}^{\alpha}$:
 $\mathcal H_{\rm e}=\mathcal H_{\rm cm}+\mathcal H_{\rm er}$, with
\begin{align}\label{}
 \mathcal H_{\rm cm}^{\alpha}&=v_{\rm F}(\varPi^{\alpha}_x\sigma_{\alpha{\rm c}}^y+c_\alpha \varPi^{\alpha}_y\sigma_{\alpha{\rm c}}^x)
 +N_{\alpha}e{\bm E}\cdot \bm R_{\alpha} ,\\
\mathcal H_{\rm er}^{\alpha}&=\sum_{j\in \alpha}\left[v_{\rm F}(\pi_j'^{x}\sigma_j^y+c_\alpha \pi_j'^{y}\sigma_j^x)
\right].
\end{align}
In this, ${\bm \varPi}^{\alpha}\equiv \bm P_{\alpha}+N_{\alpha}e\bm A(\bm R_{\alpha})=(\varPi_x^{\alpha}, \varPi_y^{\alpha})$
 is the center-of-mass canonical momentum of the $\alpha$ valley and
${\bm \pi}_j' \equiv \bm p_j'+e\bm A(\bm r_j')=(\pi_j'^x, \pi_j'^y)$
is the canonical momentum for the $j$th relative carrier. Here we
have also introduced c.m. spin operators ${\sigma}_{\alpha\rm c}^x
\equiv N_{\alpha}^{-1}\sum_{j\in \alpha}\sigma_j^x$ and
${\sigma}_{\alpha\rm c}^y \equiv N_{\alpha}^{-1}\sum_{j\in \alpha}
\sigma_j^y$ for the $\alpha$ valley. The commutation relations
between the c.m. Pauli operators $\sigma_{\alpha \rm c}^x$ and
$\sigma_{\alpha \rm c}^y$ and the Pauli operators $\sigma_j^x$ and
$\sigma_j^y$ of the $j$th carrier  are of order of $1/N_{\alpha}$.
 Therefore, for a macroscopically
large $N_{\alpha}$ system, the c.m. part $\mathcal H_{\rm cm}$
actually commutes with the relative-carrier part $\mathcal H_{\rm
er}$ in the Hamiltonian, i.e. the c.m. motion and the relative
motion of carriers are truly separated from each other. The
couplings between the two emerge only through the carrier--impurity
and carrier--phonon interactions. Furthermore, the electric field
${\bm E}$ shows up only in $\mathcal H_{\rm cm}$. And, in view of
$[r'_{i\beta_1},p'_{j\beta_2}]={\rm
i}\delta_{\beta_1\beta_2}(\delta_{ij}-1/N_\alpha)\simeq {\rm
i}\delta_{\beta_1\beta_2}\delta_{ij}$, i.e. the relative-carrier
momenta and coordinates can be treated as canonical conjugate
variables, the relative-motion part $\mathcal H_{\rm er}^{\alpha}$
is just the Hamiltonian of $N_\alpha$ carriers in the $\alpha$
valley of graphene in the magnetic field without electric field.

In terms of the c.m. coordinate ${\bm R}_{\alpha}$ and the relative
carrier density operator $\rho_{\bm q}^\alpha= \sum_{j\in \alpha}
{e}^{{\rm i}\,{\bm q}\cdot{\bm r}'_j}$, the carrier--impurity and
carrier--phonon interactions can be written
as\cite{lei1985tdb,cai1985}
\begin{align}
\mathcal H_{\rm ei}=&\sum_{\alpha,{\bm q}, a}U({\bm q})\,
{e}^{{i}\,{\bm q}\cdot \left({\bm R}_\alpha-{\bm
r}_{a}\right)}\rho_{\bm q}^\alpha,\label{1-14}\\
\mathcal H_{\rm ep}=&\sum_{\alpha,{\bm q},\nu} M({\bm q},\nu)\,
\varphi_{{\bm q}\nu}{e}^{{i}\,{\bm q}\cdot{\bm R}_\alpha}\rho_{\bm
q}^\alpha.
\end{align}
Here $U({\bm q})$  and $M({\bm q},\nu)$ are, respectively, the
impurity potential (an impurity at randomly distributed position
${\bm r}_a$) and carrier--phonon coupling matrix element in the
plane-wave representation, and $\varphi_{{\bm q}\nu}\equiv b_{{\bm
q}\nu}^{}+b_{-{\bm q}\nu}^{\dagger}$ is the phonon field operator
with $b_{{\bm q}\nu}^{\dagger}$
 and $b_{{\bm q}\nu}^{}$ being the creation and annihilation operators for a 2D phonon of wavevector
${\bm q}$ in the branch $\nu$ having frequency ${\it \Omega}_{{\bm q}\nu}$.

The derivation of balance equations starts by noticing that the c.m.
velocity (operator) ${\bm V}_{\alpha}$ is the time variation of its
coordinate: ${\bm V}_\alpha=\dot{\bm R}_\alpha=-{i}[{\bm R}_\alpha,
\mathcal H] =v_{\rm F}(\sigma_{\alpha\rm c}^y\,
\hat{i}+c_\alpha\sigma_{\alpha\rm c}^x\, \hat{j})$, and proceeds
from the Heisenberg equations for the rate of change of the
center-of-mass canonical momentum $\dot{\bm \varPi}_\alpha=-i[\bm
\varPi_\alpha,\mathcal H]$, and that of the relative-carrier energy
$\dot{\mathcal H}_{\rm er}^\alpha=-i[\mathcal H_{\rm
er}^\alpha,\mathcal H]$. The statistical average of the above
operator equations  can be determined to linear order in the
carrier--impurity and carrier--phonon interactions $\mathcal H_{\rm
ei}$ and $\mathcal H_{\rm ep}$ using the initial density matrix
$\hat \rho_0=Z^{-1}e^{-\mathcal H_{\rm ph}/T}e^{-\mathcal H_{\rm
er}/T_{\rm e}}$ with lattice temperature $T$ and a common carrier
temperature $T_{\rm e}$ for carriers in both valleys in view of the
symmetry of graphene band structure, which give rise to equal
carrier number densities $N_{K}=N_{K'}$ and equal average c.m.
velocities ${\bm v}={\bm v}_\alpha=\langle{\bm V}_\alpha\rangle$
($\alpha=K,K'$).

Steady-state transport balance equations are obtained by setting
$\sum_{\alpha}\langle \dot{\bm \varPi}_\alpha\rangle=0$ and
$\sum_{\alpha}\langle\dot{\mathcal H}_{\rm er}^\alpha\rangle=0$.
The derived force and energy balance equations, which are identical for both valleys,
can be written (for graphene system of unit area) as
\begin{align}\label{}
0=&-Ne\bm v\times\bm B-Ne\bm E+\bm f_{\rm i}+\bm f_{\rm p},\label{forceEq}\\
0=&(\bm f_{\rm i}+\bm f_{\rm p})\cdot\bm v+w.\label{energyEq}
\end{align}
Derivation of energy-balance equation is given in appendix. Here
$N=\sum_\alpha N_\alpha$ is the total number density of carriers (in
both valleys) for system of unity area, ${\bm f}_{\rm i}$ and ${\bm
f}_{\rm p}$ are total frictional forces experienced by the center of
mass due to impurity and phonon scatterings:
\begin{align}\label{}
\bm f_{\rm i}=&n_{\rm i}\sum_{\bm q}\left|U(\bm q)\right|^2\bm q\varPi_2(\bm q,\omega_0),\label{fim}\\
\bm f_{\rm p}=&\sum_{\bm q,\nu}\left|M(\bm q,\nu)\right|^2\bm q\varPi_2(\bm q,{\it \Omega}_{\bm q\nu}+\omega_0)\nonumber\\
&\hspace{0.4cm}\times\left[n\Big(\frac{{\it \Omega}_{\bm q\nu}}{T}\Big)-n\Big(\frac{{\it \Omega}_{\bm q\nu}+\omega_0}{T_{\rm e}}\Big)\right],\label{fph}
\end{align}
and $w$ is the rate of carrier energy-dissipation to the lattice due
to carrier--phonon interactions:
\begin{align}\label{}
w=&\sum_{\bm q,\nu}\left|M(\bm q,\nu)\right|^2{\it \Omega}_{\bm q\nu}\varPi_2(\bm q,{\it \Omega}_{\bm q\nu}+\omega_0)\nonumber\\
&\hspace{0.4cm}\times\left[n\Big(\frac{{\it \Omega}_{\bm q\nu}}{T}\Big)-n\Big(\frac{{\it \Omega}_{\bm q\nu}+\omega_0}{T_{\rm e}}\Big)\right].\label{fw}
\end{align}
In these equations $n_{\rm i}$ is the impurity density,
$n(x)=(e^x-1)^{-1}$ is the Bose distribution function,
$\omega_0\equiv {\bm q}\cdot {\bm v}$, $\varPi_2(\bm
q,\omega)=\sum_{\alpha}\varPi_2^{\alpha}(\bm q,\omega)$ with
$\varPi_2^{\alpha}(\bm q,\omega)$ standing for the imaginary part of
the Fourier spectrum of the relative-carrier density correlation
function of the $\alpha$ valley in the magnetic field defined by
\begin{equation}
\varPi^{\alpha}({\bm q}, t-t^{\prime})=-{i\,}\theta(t-t^{\prime})
\big\langle\big[\rho_{\bm q}^{\alpha}(t),\, \rho_{-{\bm
q}}^{\alpha}(t^{\prime})\big]\big\rangle_{0},
\end{equation}
where $\rho_{\bm q}^{\alpha}(t)={\rm e}^{{i\,}\mathcal H_{\rm
er}t}\rho_{\bm q}^{\alpha}\,{\rm e}^{-{i\,}\mathcal H_{\rm er}t}$
and $\langle...\rangle_{0}$ denotes the statistical averaging over
the initial density matrix $\hat \rho_0$.\cite{lei1985gsf,lei2008balance}

In the magnetic field the imaginary part of the relative-carrier
density correlation function $\varPi_2(\bm q,\omega)$ can be
calculated in the Landau representation.\cite{ting1977theory} The
eigenstates of the single-particle Hamiltonian $h^{\alpha}=v_{\rm
F}(\pi^x\sigma^x+c_{\alpha}\pi^y\sigma^y)$ in the magnetic field
$B\hat{z}$ can be specified by a set of quantum numbers
$\{n,k_x,\sigma,\lambda,\alpha\}$ with $n$, $k_x$, $\sigma$, and
$\lambda$ denoting the Landau index, the $x$ component of the wave
vector, the pseudospin index, and the band index (electron
$\lambda=1$ or hole $\lambda=-1$), respectively. The eigenenergies
of $h^{\alpha}$ are
\begin{equation}\label{}
\varepsilon_{\lambda n}=\lambda v_{\rm
F}\sqrt{2eBn}=\lambda\varepsilon_{n}\,\,\, (n=0,1,2,...),
\end{equation}
which is pseudospin- and valley-degenerate. The corresponding
eigenfunctions can be written as
$\Psi_{nk_x\sigma}^{\alpha\lambda}=\psi_{nk_x}^{\alpha\lambda}\otimes\chi_\sigma$
with $\chi_\sigma$ standing for the eigenstate of Pauli matrix
$\sigma_z$ and
\begin{equation}\label{}
\psi_{nk_x}^{K\lambda}(\bm r)=\frac{e^{ik_xx}}{\sqrt{1+s_n}}\left(
                                                    \begin{array}{c}
                                                      -\lambda s_{n}\phi_{{n-1},k_x}(y) \\
                                                      \phi_{n, k_x}(y) \\
                                                    \end{array}
                                                  \right),
\end{equation}
\begin{equation}\label{}
\psi_{nk_x}^{K'\lambda}(\bm r)=\frac{e^{ik_xx}}{\sqrt{1+s_n}}\left(
                                                    \begin{array}{c}
                                                      \phi_{{n},k_x}(y) \\
                                                      -\lambda s_{n}\phi_{{n-1},k_x}(y) \\
                                                    \end{array}
                                                  \right).
\end{equation}
Here $s_n=1-\delta_{n,0}$ and $\phi_{n,k_x}(y)$ is the harmonic oscillator eigenfunction:
\begin{equation}\label{}
\phi_{n,k_x}(y)=\frac{1}{\sqrt{2^nn!l_{\rm
B}\sqrt{\pi}}}\exp\left[-\frac{(y-y_c)^2}{2l_{\rm
B}^2}\right]H_n\left(\frac{y-y_c}{l_{\rm B}}\right),
\end{equation}
with $H_n(x)$ the Hermite polynomial, $l_{\rm
B}=\sqrt{1/(eB)}$ and $y_c=k_x/(eB)$.

The $\varPi_2(\bm q,\omega)$ can be expressed in the Landau-representation in the form\cite{wang2012linear,roldan2009,Pyatkovskiy}
\begin{equation}\label{piqw}
\varPi_2(\bm q,\omega)\!=\!\frac{g_sg_v}{2\pi l_{\rm
B}^2}\sum_{\substack{n,n'\\\lambda,\lambda'}} C_{n,n'}^{\lambda,\lambda'}\big(\frac{l_{\rm B}^2q^2}{2}\big)
\varPi_2(n,n';\lambda,\lambda';\omega),
\end{equation}
\begin{align}\label{p2nn}
\varPi_2(n,n';\lambda,\lambda';\omega)=&-\frac{1}{\pi}\int d\epsilon[f(\epsilon)-f(\epsilon+\omega)]\nonumber\\
&\hspace{.2cm}\times{\rm
Im}G_{\lambda n}(\epsilon+\omega){\rm Im}G_{\lambda'n'}(\epsilon).
\end{align}
Note that despite different forms of wave functions the
$\varPi_2(n,n';\lambda,\lambda';\omega)$ function and the transform
factor $C_{n,n'}^{\lambda,\lambda'}(l_{\rm B}^2q^2/2)$ are identical
for both valleys and for both pseudospin directions, whence the
valley and spin summations just give rise to the multiplication of
degenerate constants $g_v=g_s=2$. Here the transform factor
\begin{align}
C_{n,n'}^{\lambda,\lambda'}(x)= & \frac{x^{n_2-n_1}e^{-x}}{(1+s_n)(1+s_{n'})}\frac{n_1!}{n_2!}\bigg[L_{n_1}^{n_2-n_1}(x)\nonumber\\
&\hspace{1.0cm}+\lambda\lambda's_ns_{n'}\sqrt{\frac{n_2}{n_1}}L_{n_1-1}^{n_2-n_1}(x)\bigg]^2,
\end{align}
 with
$n_1={\rm min}(n,n')$, $n_2={\rm max}(n,n')$, and $L_n^m(x)$ being associated Laguerre polynomials.

The Landau levels are broadened due to impurity, phonon and
carrier-carrier scatterings. We model the imaginary part of the
retarded Green's function ${\rm Im}G_{\lambda n}(\epsilon)$ in
Eq.\,(\ref{p2nn}), or the density-of-state (DOS) of the $\lambda
n$th Landau level, using a Gaussian form\cite{Ando1982}
\begin{equation}\label{}
{\rm Im}G_{\lambda
n}(\epsilon)=-\frac{\sqrt{2\pi}}{\varGamma_{\lambda n}}
\exp\left[-\frac{2(\epsilon-\varepsilon_{\lambda n})^2}
{\varGamma_{\lambda n}^2}\right],
\end{equation}
with a half-width\cite{Zheng2002}
\begin{equation}
\varGamma_{\lambda n}=\left[{2\omega_{\lambda
n}}/({\pi\tau_s})\right]^{1/2},
\end{equation}
 where $\tau_s$ is the single-particle lifetime and
 $\omega_{\lambda n}=|\varepsilon_{\lambda n+1}-\varepsilon_{\lambda n}|$ is the level distance or the cyclotron frequency
 of the $\lambda n$th Landau level, with $\omega_{\lambda n}\approx
 v_{\rm F}(eB/2n)^{1/2}=eBv_{\rm F}^2/\varepsilon_n$ for large $n$ irrespective of the
 band index, giving rise to valley- and band-independent broadening $\varGamma_{\lambda n}=\varGamma_{n}$.

In the following we restrict ourselves to the $n$-doped
case at relatively low temperature, i.e., the carriers are electrons,
that we only need to consider states with band index $\lambda=1$.
For conciseness we will no longer write out the band index $\lambda$ in the expressions and equations
and denote $\varPi_2(n,n';1,1;\omega)$,
$C_{n,n'}^{1,1}(x)$, and ${\rm Im}G_{1 n}(\epsilon)$  simply as
$\varPi_2(n,n';\omega)$, $C_{n,n'}(x)$ and ${\rm Im}G_{n}(\epsilon)$.
The Landau-level summation indices $n$ and $n'$ in all the equations are taken over $0,1,2,...$ but
the ${\rm Im}G_{0}(\epsilon)$ function should be replaced by ${\rm Im}G_{0}^{p}(\epsilon)=\theta(\epsilon){\rm
Im}G_{0}(\epsilon)$ due to electron-hole symmetry of the band structure.\cite{wang2012linear}

The total electron number density $N$ is related to the chemical potential $\varepsilon_f$
of the Landau quantized graphene system by the equation
\begin{equation}\label{den}
N=-\frac{g_sg_v}{2(\pi l_{\rm
B})^2} \sum_n\int d\epsilon
f(\epsilon) {\rm Im}G_n(\epsilon),
\end{equation}
in which $f(\epsilon)=\{\exp[(\epsilon-\varepsilon_{f})/T_{\rm
e}]+1\}^{-1}$ is the Fermi distribution function at electron temperature $T_{\rm e}$.

Force- and energy-balance equations (\ref{forceEq}) and
(\ref{energyEq}), in which the frictional forces ${\bm f}_{\rm i}$,
${\bm f}_{\rm p}$ and the electron dissipation rate $w$ are
functions of carrier drift velocity ${\bm v}$ and electron
temperature $T_{\rm e}$, describe the steady-state nonlinear
magnetotransport in the graphene. With given carrier drift velocity
${\bm v}$ or the dc current density ${\bm J}=Ne{\bm v}$, the
electron temperature $T_{\rm e}$ can be determined by the
energy-balance equation, and the magnetoresistance is obtained from
force-balance equation. Note that the frictional forces $\bm f_{\rm
i}$ and  $\bm f_{\rm p}$ are in the opposite direction of the drift
velocity $\bm v$ and their magnitudes are functions of $v=|{\bm v}|$
only: ${\bm f}_{\rm i}=-{\bm v}f_{\rm i}(v)/v$ and ${\bm f}_{\rm
p}=-{\bm v}f_{\rm p}(v)/v$. In the Hall configuration, e.g. with a
drift velocity $\bm v=(v,0)$ in the $x$ direction, the force-balance
equation Eq.\,\eqref{forceEq} yields a transverse resistivity
$R_{xy}=-E_y/(Nev)=-B/(Ne)$,  a longitudinal resistivity
$R_{xx}=-E_x/(Nev)=-(f_{\rm i}+f_{\rm p})/(N^2e^2v)$, and a
longitudinal differential resistivity
$r_{xx}=-(N^{2}e^{2})^{-1}d(f_{\rm i}+f_{\rm p})/d v$.

\section{Numerical calculations and discussions}

We will use a phenomenological parameter $\alpha_{\Gamma}$ to relate
the single particle lifetime $\tau_s$ to the transport scattering
time in the system:\cite{lei2003radiation} $\tau_{\rm
tr}=\alpha_{\Gamma}\tau_{s}$, and, by expressing $\tau_{\rm tr}$
with the zero-field mobility $\mu$,\cite{hwang2007,wang2011}
 we can write the Landau-level broadening in the vicinity of Fermi energy $\varepsilon_{\rm F}=v_{\rm F}\sqrt{\pi N}$ as
\begin{equation}
\varGamma=(ev_{\rm F}/\pi)[2B\alpha_{\Gamma}/(N\mu)]^{1/2}.
\end{equation}
The broadening parameter will be taken to be $\alpha_{\Gamma}=2$
throughout the calculation.

We consider two cases: a graphene monolayer on a SiO$_2$ substrate\cite{tan2011shubnikov} and a suspended monolayer graphene.
The electrons in graphene are scattered by charged
 impurities distributed at a distance $d$ from the layer with $d=4\,{\rm \AA}$ for the graphene on SiO$_2$ substrate
 and $d=0$ for the suspended one, having a scattering potential
\begin{equation}
U(\bm q)=\frac{Ze^2}{2\epsilon_0\kappa_{\rm avg}q}e^{-qd}. \label{coulomb}
\end{equation}
Here $\kappa_{\rm avg}$ is the average dielectric constant of two
regions (air and SiO$_2$ or air) surrounding the graphene.
Hence\cite{aniruddha2010effect,Fischetti} $\kappa_{\rm
avg}\approx (1+\kappa)/2=2.45$ for non-suspended graphene ($\kappa=3.9$ is the
static dielectric constant of SiO$_2$), while  $\kappa_{\rm
avg}\approx 1$ for suspended one.

For intrinsic acoustic phonon scatterings in the graphene layer, there are two
2D modes, the sum of which
can be treated as isotropic one\cite{Vasili2010inelastic,Raseong2011}
with a scattering matrix element
\begin{equation}
|M(\bm q,{\rm AC})|^2=\frac{D^2q}{2\rho_mv_{\rm ph}},
\end{equation}
and an averaged sound velocity\cite{Hwang2008acoustic} $v_{\rm
ph}=2\times10^4\,{\rm m/s}$.
We choose the deformation potential constant as a
moderate value\cite{Hwang2008acoustic,chen2008intrinsic}
$D=19\,{\rm eV}$ and the mass density $\rho_m=7.6\times10^{-8}\,{\rm g/cm^2}$.\cite{Hwang2008acoustic}

The electrons can also be scattered by the intrinsic optical phonons
in graphene. However, the energies of these intrinsic optical modes
are greater than $150\,{\rm meV}(\approx1740\,{\rm K})$, which is
much larger than the lattice and electron temperatures concerned and
can be neglected. For graphene on the SiO$_2$ substrate, the surface
optic phonon couples to the electrons in graphene by an effective
electric field. Due to small van der Waals distance between the
polar substrate and the interface, the 2D surface optical (SO)
phonon plays a more prominent role in transport in graphene than in
usual heterojunctions. The coupling matrix element can be written
as\cite{fratini2008substrate}
\begin{equation}
|M(\bm q,{\rm SO})|^2=\frac{e^2 \varOmega_{\rm so}}{2\epsilon_0 q}
\left(\frac{1}{\kappa_\infty+1}-\frac{1}{\kappa+1}\right)e^{-2qd},
\end{equation}
where $\varOmega_{\rm so}$ is the frequency of SO phonon and $\kappa_\infty$ is the optical dielectric constant of
substrate. For SiO$_2$, $\kappa_\infty=2.4$ and there are two
SO-phonon modes having frequencies\cite{Fischetti} $\varOmega_{\rm
so}^{(1)}=59\,{\rm meV}$ and $\varOmega_{\rm so}^{(2)}=155\,{\rm
meV}$. The second mode is negligible in the present study owing to
its large frequency.

\subsection{SdHO under nonzero dc current}\label{sdho}

In order to study the SdHO under a finite bias dc current in
graphene we calculate the magnetoresistivity of a graphene monolayer
on a SiO$_2$ substrate having electron density
$N=3.16\times10^{12}\,{\rm cm^{-2}}$ and zero-magnetic-field
mobility $\mu=0.8\,{\rm m^2/Vs}$ in the magnetic fields  ranging
from 0 to 15\,T at lattice temperature $T=2\,{\rm K}$ on the basis
of balance equations (\ref{forceEq}) and (\ref{energyEq}). The
calculated longitudinal magnetoresistivity $R_{xx}$ and differential
magnetoresistivity $r_{xx}$ are shown in Fig.\,1(a) and Fig.\,1(b)
as functions of the magnetic field $B$ for different given current
densities $J$. The standard SdHO curves of graphene are obtained,
where the valleys of magnetoresistivity $R_{xx}$ locate at the
magnetic fields corresponding to the half-integer filling
factors\cite{zhang2005experimental,tan2011shubnikov}
$\nu=\tfrac{2\pi N}{eB}=4(n+\tfrac{1}{2})$ with $n=2,3,4,...,$ as
indicated in the figure. The increasing current density suppresses
the oscillation, while the peak/valley positions remain essentially
unchanged. The significant feature of current-related SdHO appears
in the differential resistivity as shown in Fig.\,1(b). With the
rise of current density, the oscillation of differential resistivity
$r_{xx}$ not only tends to decrease its amplitude, but, more
prominently, exhibits phase inversion, e.g., SdHO minima (maxima)
invert to maxima (minima) at certain value of bias current density,
which is roughly linearly dependent on the magnetic field of the
SdHO extrema. These features are in good agreement with the
experimental observation.\cite{tan2011shubnikov}

\begin{figure}
\begin{center}
  \includegraphics[width=0.42\textwidth]{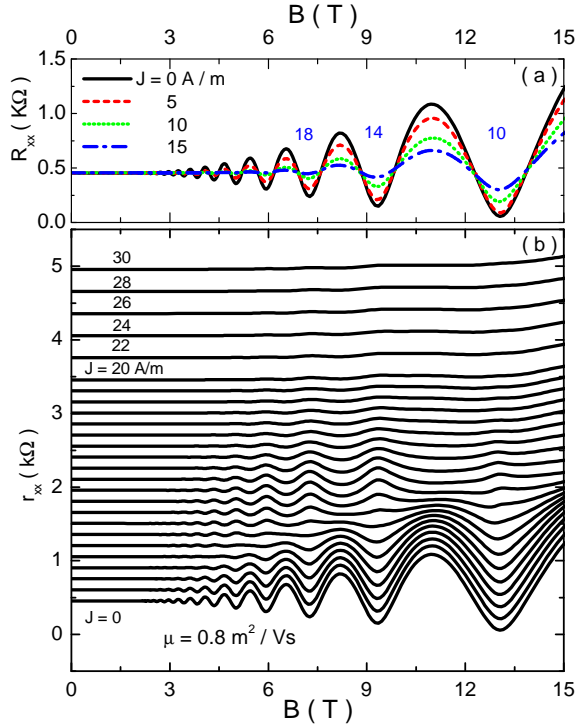}
\end{center}
\caption{(Color online) (a) Magnetoresistivity $R_{xx}$ is shown versus magnetic field $B$
at various dc current densities $J=$0, 5, 10, 15$\,{\rm A/m}$. The integers near
the valleys indicate the filling factors. (b) Differential magnetoresistivity
$r_{xx}$ is plotted as a function of the magnetic field for
various current densities at lattice temperature $T=2\,{\rm K}$.
These $r_{xx}$ curves of different $J$ values are vertically offset for clarity.
The current densities are $J=0$, 1, 2, $\cdots$,
20$\,{\rm A/m}$ in 1$\,{\rm A/m}$ step for the lower 21 ones, or
are indicated in the figure for others. The system is a monolayer graphene on a SiO$_2$ substrate
having electron density $N=3.16\times10^{12}\,{\rm cm^{-2}}$ and zero-magnetic-field
mobility $\mu=0.8\,{\rm m^2/Vs}$.}\label{with}
\end{figure}

\begin{figure}
\begin{center}
  \includegraphics[width=0.40\textwidth]{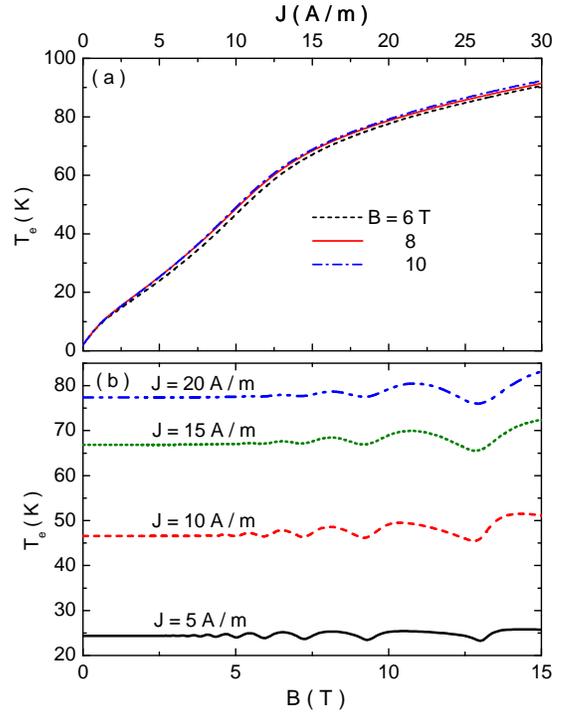}
\end{center}
\caption{(Color online) Electron temperature $T_{\rm e}$ is shown as
a function of dc bias current density at various magnetic fields (a)
and as a function of magnetic field at various bias current
densities (b) for the same system as described in Fig.\,1.}
\label{Te}
\end{figure}

The phase inversion of SdHO is closely related to the rise of electron
temperature with increasing bias current. Fig.\,\ref{Te} shows the calculated
electron temperature $T_{\rm e}$ as a function of the bias current
density $J$ at magnetic field strengths $B=6,8$, and $10$\,T (a), as
well as $T_{\rm e}$ versus $B$ at current densities
$J=5,10,15,20$\,A/m (b). When current density is lower than 12\,A/m,
the electron temperature almost linearly depends on the dc bias. For
higher current density, the enhanced energy dissipation arising from
electron--SO-phonon interaction restrains the linear increase of
electron temperature. In the fixed bias current case (b), only a small oscillation
of electron temperature around a certain value shows up for almost the whole
magnetic field range presented in the figure.

\begin{figure}
\begin{center}
  \includegraphics[width=0.48\textwidth]{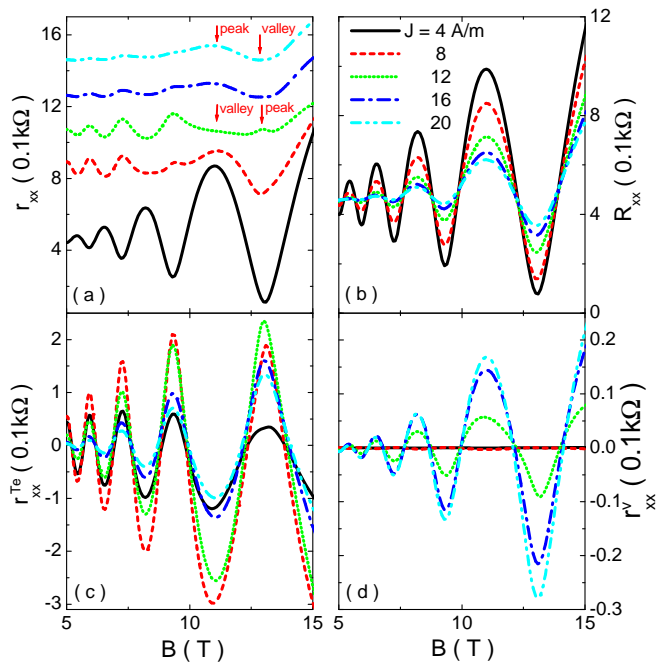}
\end{center}
\caption{(Color online) Differential magnetoresistivity $r_{xx}$ (a), and its constituent parts $R_{xx}$ (b),
$r_{xx}^{T_{\rm e}}$ (c) and $r_{xx}^{v}$ (d) defined in (\ref{components}), are shown versus magnetic field for
various bias current densities $J=4$, 8, 12, 16, 20$\,{\rm A/m}$. The $r_{xx}$ curves
in (a) are vertically offset for clarity.}\label{rte}
\end{figure}

In the balance-equation scheme, the frictional forces $f_{\rm i}$ and $f_{\rm p}$ are functions of the
drift velocity $v$ (i.e. the current density $J=Nev$) and the electron temperature $T_{\rm e}$, and the latter
is determined as a function of $v$ from the energy balance equation. Therefore the differential
resistivity derived can be expressed as
\begin{align}
r_{xx}&=R_{xx}+J\frac{\partial R_{xx}}{\partial T_{\rm e}}\frac{\partial T_{\rm e}}{\partial J}+J\frac{\partial R_{xx}}
{\partial J},\nonumber\\&=R_{xx}+r_{xx}^{T_{\rm e}}+r_{xx}^v, \label{components}
\end{align}
where $r_{xx}^{T_{\rm e}}$ can be thought as the part arising from the electron-temperature change and
$r_{xx}^{v}$ as that direct from current-density change.
 We plot the calculated
 $R_{xx}$, $r_{xx}^{T_{\rm e}}$ and $r_{xx}^{v}$, as well as the total $r_{xx}$, as functions of
 the magnetic field $B$ for several bias
current densities $J=4,8,12,16$, and $20$\,A/m in Fig.\,\ref{rte}.
The three constituent parts $R_{xx}$, $r_{xx}^{T_{\rm e}}$ and $r_{xx}^{v}$ all
exhibit oscillations having extrema at positions
$\nu=4(n+\tfrac{1}{2})$. However, the phase of $r_{xx}^{T_{\rm
e}}$ is opposite to those of $R_{xx}$ and $r_{xx}^{v}$. Note that in
the current range $0<J<12\,{\rm A/m}$, when SO-phonons play a
relatively small role in dissipating energy, the electron
temperature grows almost linearly with increasing current density
and $|r_{xx}^{v}|$ is one order of magnitude smaller than $|R_{xx}|$
or $|r_{xx}^{T_{\rm e}}|$, hence, $R_{xx}$ and $r_{xx}^{T_{\rm e}}$
constitute dominant contributions to total $r_{xx}$ and the
current-induced electron temperature rising accounts for the phase
inversion of $r_{xx}$ in this current density regime, as pointed out
by Tan {\it et al.}\cite{tan2011shubnikov}

With further increase in the current density, $R_{xx}$ decreases,
while $r_{xx}^{T_{\rm e}}$ first ascends and then descends in view
of the slowdown of the electron temperature increase due to the
enhanced role of SO-phonon scattering. On the other hand, at higher
current density $J$, the current direct-contributed part, $r_{xx}^{v}$,
 also becomes non-negligible. This could
give rise to a second phase-inversion of $r_{xx}$ oscillation. It can be
seen in Fig.\,\ref{rte}(a) that the peak (valley) at low current density
near 11\,T (13\,T) first inverts to valley (peak) and then changes back
to peak (valley) with the rise of dc bias.

\subsection{Current-induced magnetoresistance oscillation}\label{CIMO}

We turn to the regime of lower magnetic fields, where the SdHO hardly shows up.

In the case of low temperature $T_{\rm e}\ll \varepsilon_{\rm F}$
and large filling factor $\nu=\tfrac{\pi N}{2eB} \gg 1$,
the major contribution to the summation
in the density correlation function (\ref{piqw}) comes from Landau
levels near the Fermi energy, i.e., terms $n\simeq n'\sim \nu$, and
then the function $C_{n,n'}(x)$ has a sharp principal maximum near
$x\sim 4\nu$. Therefore, as a function of the in-plane momentum
${q}$, the $\varPi_2(\bm q,\omega)$ function given in (\ref{piqw})
sharply peaks around $q \approx 2k_{\rm F}$, with $k_{\rm
F}=\sqrt{\pi N}$ being the Fermi wave vector. In the case of a
finite drift velocity $v$, the motion of the center-of-mass provides
the relative electron with an additional energy $\omega_0={\bm
q}\cdot{\bm v}$ during its transition from a state to another state
having a momentum change of ${\bm q}$, as shown in the expressions
of (\ref{fim}), (\ref{fph}) and (\ref{fw}) for $\bm f_{\rm i}$, $\bm
f_{\rm p}$ and $w$. The sharp peaking of $\varPi_2(\bm q,\omega)$
function around $q \approx 2k_{\rm F}$ indicates that most effective
processes contributing to the magnetoresistance come from those
electron transitions which involve an additional energy around
$\omega_j=2k_{\rm F}v$. Looking at electron transitions in the
Landau representation, we can see that the transition rate is
proportional to the overlap of the DOS of the related two Landau
levels around the Fermi surface, ${\rm
Im}G_{n}(\epsilon+\omega_j){\rm Im}G_{n'}(\epsilon)$, and the
maximum overlap occurs at $\varepsilon_{n}-\varepsilon_{n'} =
\omega_j$. Thus, the impurity-induced longitudinal
magnetoresistivity may show extrema when
$\varepsilon_{\nu+l}-\varepsilon_{\nu} \approx l\omega_{\rm B}=\pm
\omega_j$ with $l=0,\pm 1, \pm2, ...$ and $\omega_{\rm B}=eBv_{\rm
F}/k_{\rm F}$ being the distance of the neighboring Landau levels in
the vicinity of Fermi surface. Therefore, the impurity-related
magnetoresistivity would exhibits a periodical oscillation when
changing drift velocity $v$ or changing magnetic field $B$. This
current-induced magnetoresistance oscillation (CIMO) is
characterized by a dimensionless parameter $\omega_j/\omega_{\rm B}$
with a period $\Delta (\omega_j/\omega_{\rm B})\approx 1$: when
$\omega_j/\omega_{\rm B}$ varies by a unity value, the
magnetoresistivity experiences change of an oscillatory period.

\begin{figure}
\begin{center}
  \includegraphics[width=0.45\textwidth]{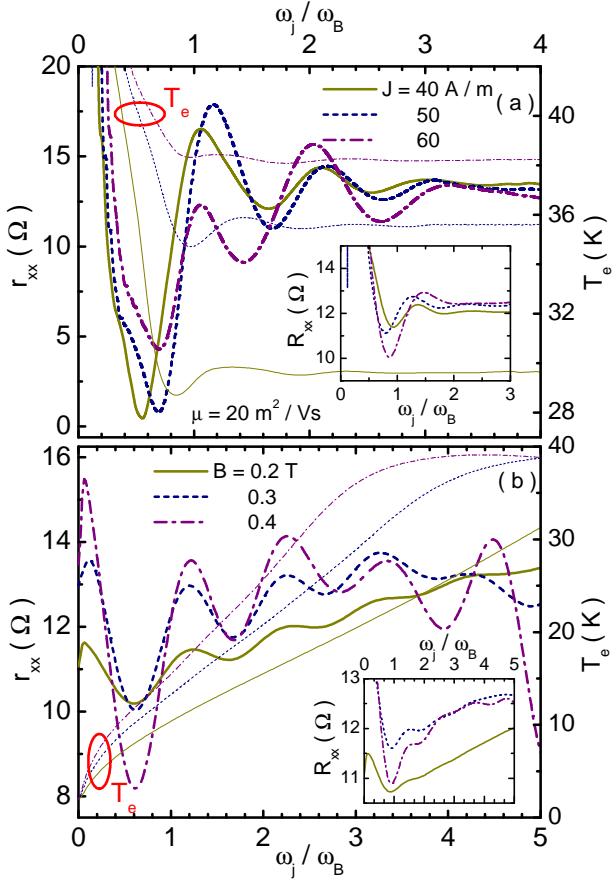}
\end{center}
\caption{(Color online) Differential magnetoresistivity $r_{xx}$, electron temperature $T_{\rm e}$
and magnetoresistivity $R_{xx}$ (inset)
are shown as functions of $\omega_j/\omega_{\rm B}$ for various fixed dc current densities $J=$40, 50,
60$\,{\rm A/m}$ (a) and for various fixed magnetic fields $B=0.2,
0.3, 0.4\, {\rm T}$ (b).  The system is a suspended monolayer graphene having
electron density $N=3.16\times10^{12}\,{\rm cm^{-2}}$ and zero-magnetic-field mobility $\mu=20\,{\rm m^2/Vs}$
at lattice temperature $T=2\,{\rm K}$, with Coulombic impurity potential (\ref{coulomb}) of $d=0$.}\label{cio}
\end{figure}

As an example, Fig.\,\ref{cio} displays the calculated magnetoresistivity
and differential magnetoresistivity versus $\omega_j/\omega_{\rm B}$
for fixed bias current densities $J=40, 50$ and $60$\,A/m (a) or
for fixed magnetic fields $B=0.2, 0.3$ and $0.4$\,T (b)
at lattice temperature $T=2$\,K
in a suspended monolayer graphene having electron density
$N=3.16\times 10^{12}\,{\rm cm^{-2}}$ and linear mobility $\mu=20\,{\rm m^2/Vs}$,
assuming Coulombic impurity scattering potential (\ref{coulomb}) with $d=0$.
The longitudinal magnetoresistivity  $R_{xx}$ (plotted in the insets)
shows relatively weak oscillations,
while the differential magnetoresistivity $r_{xx}$ exhibits marked oscillations,
having an approximate period $\Delta(\omega_j/\omega_{\rm B})\approx 1$ in both cases.
Notable magnetoresistance oscillations appear in the well-resolved Landau level regime
when $2\varGamma\leq \omega_{\rm B}$, or $B\geq 8\alpha_{\Gamma}/\pi\mu\approx 0.25\,$T,
and the enhanced current weakens the oscillation amplitude due to the
rising electron temperature.

Note that the parameter $\omega_j/\omega_{\rm B}=(2\pi/e^2 v_{\rm
F})(J/B)$ characterizing the CIMO depends only on the
band-dispersion related $v_{\rm F}$ for systems of linear energy
band, thus the periodic behavior of CIMO is universal in graphene in
terms of $J/B$, irrespective of carrier-density $N$. This situation
is in contrast to the conventional 2DEG of parabolic
band,\cite{lei2007current} where the Fermi velocity $v_{\rm F}$
involved in the characterizing parameter depends on the carrier
density, so does the periodicity of the magnetoresistance
oscillation in it.

The basic features of the oscillatory $R_{xx}$ and $r_{xx}$ are: oscillation
amplitude decays with increasing $\omega_j/\omega_{\rm B}$ but enhances
with increasing current density or magnetic field strength in the discussed range.
In the fixed current density case of Fig.\,\ref{cio}(a),
where the electron temperature has only weak change
with changing magnetic field, the amplitude decrease of the resistance oscillation
is due to the enlarged overlap of neighboring Landau levels
with decreasing magnetic field.
In the fixed $B$-field case of Fig.\,\ref{cio}(b), the electron temperature grows
when increasing bias current density,
resulting in the suppression of the resistance oscillation.
Nevertheless, the oscillation amplitude shown in these figures exhibits somewhat anomalous behavior,
especially around the first peak of $J=60\,{\rm A/m}$ curve in Fig.\,\ref{cio}(a)
and the last peak of $B=0.4\,{\rm T}$ curve in Fig.\,\ref{cio}(b).
These $r_{xx}$ anomalies come from the contribution of
phonon-related differential resistivity $r_{\rm ph}$.

\begin{figure}
\begin{center}
  \includegraphics[width=0.48\textwidth]{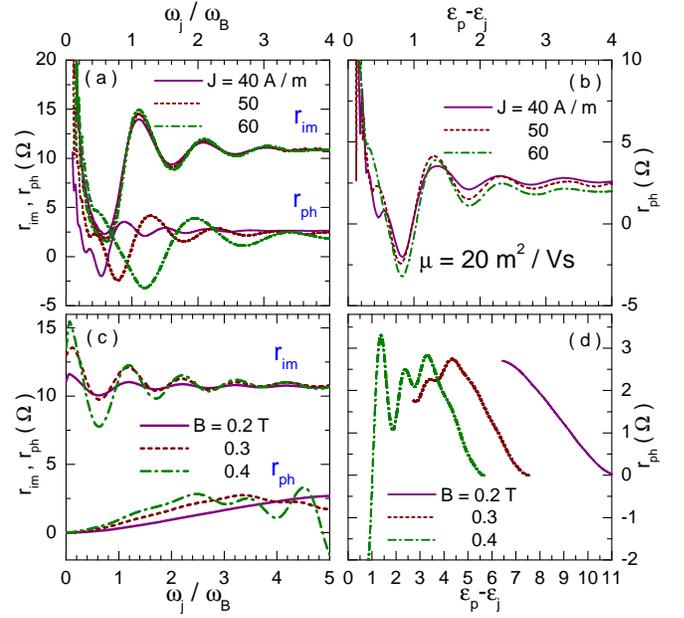}
\end{center}
\caption{(Color online) Impurity-related and phonon-related
differential resistivities $r_{\rm im}$ and $r_{\rm ph}$ are shown versus
$\omega_j/\omega_{\rm B}$ for fixed current densities $J=40, 50, 60\,{\rm
A/m}$ (a) or fixed magnetic fields $B=0.2, 0.3, 0.4\, {\rm T}$ (c).
Phonon-related differential resistivity $r_{\rm ph}$ is replotted as a function
of $\varepsilon_p-\varepsilon_j$ for fixed current densities (b) or fixed
magnetic fields (d).}\label{ciodif}
\end{figure}

In contrast to the case of high-mobility 2DEG,\cite{lei2007current}
the electron temperature $T_{\rm e}$ in the present monolayer
graphene may reach the range of 40\,K in the case of high current
density $J\geq 60$\,A/m and the magnitude of phonon-related
resistivity may not be negligible in comparison with impurity
contribution as shown in Fig.\,\ref{ciodif}(a) and (c), where the
constituent parts of $r_{xx}$ in the monolayer graphene, the
resistivity  $r_{\rm im}=-(N^{2}e^{2})^{-1}df_{\rm i}/d v$ due to
impurity scattering and the resistivity $r_{\rm
ph}=-(N^{2}e^{2})^{-1}df_{\rm p}/d v$ due to intrinsic acoustic
phonon scattering, are plotted as functions of $\omega_j/\omega_{\rm
B}$ respectively for the cases of fixed current density (a) and for
the cases of fixed magnetic field strength (c). The oscillation
behavior of $r_{\rm im}$ closely follows the basic feature of CIMO,
but $r_{\rm ph}$, though generally smaller in magnitude, appears
quite different. In the fixed current case the marked drop of
$r_{\rm ph}$ around $\omega_j/\omega_{\rm B} \sim 1$
[Fig.\,\ref{ciodif}(a)] leads to the descent of the first peak of
$r_{xx}$ at $J=60\,{\rm A/m}$ curve in Fig.\,\ref{cio}(a). In the
fixed magnetic field case, the resonant peak of $r_{\rm ph}$ around
$\omega_j/\omega_{\rm B} \sim 4.5$ for $B=0.4\,{\rm T}$
[Fig.\,\ref{ciodif}(c)] gives rise to the enhancement and position
shift of the last peak of $r_{xx}$ in Fig.\,\ref{cio}(b).

Such kind of oscillatory $r_{\rm ph}$ is referred to the magnetophonon resonance induced by acoustic phonons.
As in conventional 2DEGs,\cite{Zudov2001new,zhang2008resonant,lei2008low} acoustic phonon-related resistivity $r_{\rm ph}$
in a dc biased graphene should feature a periodical appearance of resonant peaks
with respect to $\varepsilon_p-\varepsilon_j$ axis,
where $\varepsilon_j \equiv \omega_j/\omega_{\rm B}$ and $\varepsilon_p \equiv \omega_{\rm ph}/\omega_{\rm B}$
are the ratios
of the energy $\omega_j$ provided by the drifting center-of-mass and the energy $\omega_{\rm ph}=2k_{\rm F}v_{\rm ph}$
provided by the optimum phonons to the inter-Landau-level distance $\omega_{\rm B}$ of electron near the Fermi surface.
We replot the phonon-related resistivities $r_{\rm ph}$ given in Fig.\,\ref{ciodif}(a) and (c)
as a function of $\varepsilon_p-\varepsilon_j$ in Fig.\,\ref{ciodif}(b) and (d).
They indeed show peaks near integer positions $\varepsilon_p-\varepsilon_j\approx l=1,2,3$ and $4$,
indicating electron scattered resonantly across $l$ Landau-level spacings
by absorbing or emitting an optimum acoustic phonon under the biased dc current condition.
At low magnetic fields, the magnetophonon resonance in $r_{\rm ph}$ can not be seen
in the range shown, because of weakened oscillation in the DOS and higher orders of resonant peaks required
(e.g., $\varepsilon_p-\varepsilon_j \geq 6$ for $0\leq\varepsilon_j\leq5$ at $B=0.2\,{\rm T}$ ).

\begin{figure}
\begin{center}
  \includegraphics[width=0.42\textwidth]{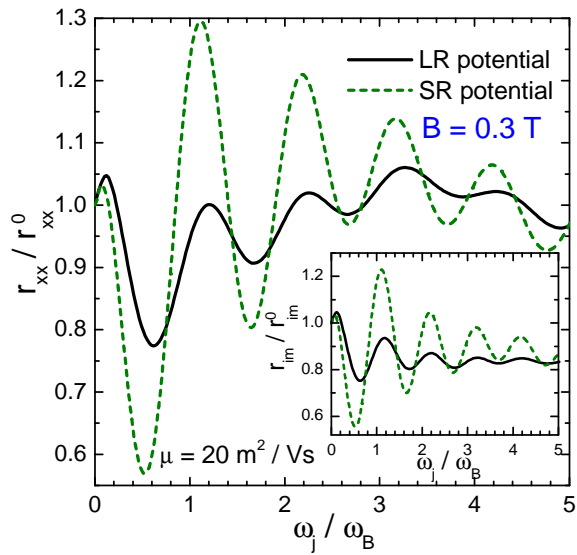}
\end{center}
\caption{(Color online) Normalized differential resistivity versus $\omega_j/\omega_{\rm B}$
at fixed magnetic field $B=0.3\,{\rm T}$ for the system subject to LR or SR impurity-scattering.
The inset displays normalized impurity-related differential resistivity $r_{\rm im}$. Here $r_{xx}^0$ and $r_{\rm im}^0$
are total and impurity-related differential resistivity in the absence of magnetic field.
The zero-magnetic-field mobility $\mu=20\,{\rm m^2/Vs}$.}\label{short}
\end{figure}

Analogous to the case of 2DEG,\cite{lei2003radiation,lei2010mater}
the amplitude of current-controlled magnetoresistance oscillation
depends strongly on the correlation length of electron-impurity
scattering potential, though the oscillation periods are essentially
the same in terms of $\varepsilon_j$. To see this, we plot the
normalized total and impurity-induced differential resistivities,
$r_{xx}$ and $r_{\rm im}$, for Coulombic impurity scattering
potential (\ref{coulomb}) with $d=0$ (LR) and short-range (SR)
disorders (assuming same zero-magnetic-field mobility $\mu=20\,{\rm
m^2/Vs}$ and $\alpha_{\Gamma}=2$ for both cases) in
Fig.\,\ref{short} as functions of $\omega_j/\omega_{\rm B}$ at fixed
magnetic field $B=0.3\,{\rm T}$. The lattice defects in graphene are
usually modeled by SR impurities. It is seen that both $r_{xx}$ and
$r_{\rm im}$ display much stronger oscillations in the case of SR
potential than that of LR potential, but the maxima and minima
positions are almost identical in both cases.

\section{summary}\label{summary}
In summary, we have presented an investigation of nonlinear
magnetotransport in graphene under a finite dc bias at low
temperature employing a balance-equation scheme appropriate to
systems with linear-energy dispersion. In the relatively strong
magnetic field range when SdHO controlled by the filling factor
$\nu=2\pi N/e B$ shows up we find that the oscillatory differential
magnetoresistivity exhibits phase inversion with rising bias current
density, in agreement with recent experimental finding. Further, it
is demonstrated that electron--SO-phonon scattering is important for
graphene on a polar substrate, which suppresses the rapid increase
of electron temperature and may result in a second phase inversion
of the oscillatory resistance. In the lower magnetic field and
higher bias current density regime when SdHO becomes weak a CIMO is
appreciable in suspended graphene. It appears markedly in the
differential resistivity when Landau levels are still well resolved
and is controlled by the parameter $\varepsilon_j=(2\pi/e^2 v_{\rm
F})(J/B)$ having approximate period $\Delta \varepsilon_j \sim 1$.
For the graphene mobility available today ($\approx 20\,{\rm
m^2/Vs}$),
 the oscillatory behavior may be some what altered by magnetophonon resonance induced
 by intrinsic acoustic phonon under finite bias current.
We hope this current-controlled magnetoresistance oscillation could
be observed experimentally in the near future.

\section*{ACKNOWLEDGMENTS}
This work was supported by the National Basic Research Program of
China (Grant No. 2012CB927403), the National Science Foundation of China
(Grant No. 11104002), the Program for Science\&Technology Innovation
Talents in Universities of Henan Province (Grant No. 2012HASTIT029), and
the Program of Young Key Teachers of University in Henan Province
(Grant No. 2011GGJS-148).

\section*{Appendix: Derivation of the energy-balance equation}\label{appendix}
Here we detail the derivation of the energy-balance equation for
graphene. In the second quantization representation of the creation
(annihilation) operators $c_{\alpha\lambda nk_xs}^{\dag}$
($c_{\alpha\lambda nk_xs}$), the relative-carrier Hamiltonian has
the form:
\begin{equation}\label{}
\mathcal H_{\rm er}=\sum_{\alpha,\lambda, n,k_x,s}\varepsilon_{\lambda n}c_{\alpha\lambda nk_xs}^{\dag}c_{\alpha\lambda nk_xs}
\end{equation}
The rate of change of the energy of relative carrier system is
obtained from the Heisenberg equation of motion:
\begin{align}\label{operator}
\dot{\mathcal H}_{\rm er}=&-i[\mathcal H_{\rm er},\mathcal H]\nonumber\\
=&-\sum_{\bm q,a}U(\bm q,z_a)e^{i\bm q\cdot(\bm R-\bm r_a)}\frac{d\rho_{\bm q}(t)}{dt}\nonumber\\
&-\sum_{\bm q,\nu}M(\bm q,\nu)e^{i\bm q\cdot\bm R}\varphi_{\bm
q\nu}(t)\frac{d\rho_{\bm q}(t)}{dt}.
\end{align}
Here the particle density operator
\begin{align}
\rho_{\bm q}(t)=\sum_{\substack{\alpha,s,\lambda,n,k_x\\ \alpha',s',\lambda',n',k_x'}}&\langle\Psi_{nk_xs}^{\alpha\lambda}
|e^{i\bm q\cdot\bm r}|\Psi_{n'k_x's'}^{\alpha'\lambda'}\rangle e^{i(\varepsilon_{\lambda n}-\varepsilon_{\lambda' n'})t}\nonumber\\
&\times c_{\alpha\lambda nk_xs}^{\dag}c_{\alpha'\lambda' n'k_x's'}
\end{align}
After statistical average of the operator equation \eqref{operator}, the energy-balance equation is given by\cite{lei2008balance}
\begin{align}
\frac{dU}{dt}=\left\langle\frac{d\mathcal{H}_{\rm er}}{dt}\right\rangle=I_1+I_2,
\end{align}
with
\begin{align}
I_1=i\int_{-\infty}^{t}dt'n_{\rm i}\sum_{\bm q}&|U(\bm q)|^2e^{i\bm q\cdot[\bm R(t)-\bm R(t')]}\nonumber\\
&\times\left\langle\left[\frac{d\rho_{\bm q}(t)}{dt},\rho_{-\bm q}(t')\right]\right\rangle_0,
\end{align}
\begin{align}
I_2=i\int_{-\infty}^{t}dt'\sum_{\bm q,\nu}&|M(\bm q,\nu)|^2e^{i\bm q\cdot[\bm R(t)-\bm R(t')]}\nonumber\\
&\times\left\langle\left[\varphi_{\bm q\nu}(t)\frac{d\rho_{\bm
q}(t)}{dt},\varphi_{-\bm q\nu}(t')\rho_{-\bm q}(t')
\right]\right\rangle_0.
\end{align}
The first integral $I_1$ can be simplified as
\begin{align}
I_1=&-\int_{-\infty}^{\infty}dt'n_{\rm i}\sum_{\bm q}|U(\bm q)|^2e^{i\bm q\cdot\bm v(t-t')}\frac{d}{dt}\varPi(\bm q, t-t')\nonumber\\
&-i\sum_{\bm q}|U(\bm q)|^2\left\langle\left[\rho_{\bm
q}(t),\rho_{-\bm q}(t)\right]\right\rangle_0.
\end{align}
Here the relative-carrier density correlation function $\varPi(\bm
q,t-t')=-i\theta(t-t') \left\langle\left[\rho_{\bm q}(t),\rho_{-\bm
q}(t')\right]\right\rangle_0$. The second term of the above equation
equals zero and the first term becomes $-\bm f_{\rm i}\cdot\bm v$
after integration by parts, hence we obtain $I_1=-\bm f_{\rm
i}\cdot\bm v$. Similarly, the integral $I_2=-\bm f_{\rm p}\cdot\bm
v-w$. Therefore, the energy-balance equation is written as
\begin{align}
\frac{dU}{dt}=\left\langle\frac{d\mathcal{H}_{\rm er}}{dt}\right\rangle=-(\bm f_{\rm i}+\bm f_{\rm p})\cdot\bm v-w.
\end{align}

\end{document}